\documentclass[aps,prl,twocolumn,nofootinbib,showpacs,psfig,superscriptaddress,longbibliography]{revtex4-1}

\usepackage{color}
\usepackage[colorlinks=true,linkcolor=blue,citecolor=blue]{hyperref}
\usepackage{hyperref}
\usepackage{amsmath}
\usepackage{amssymb}
\usepackage{graphicx}
\usepackage{epsfig}
\usepackage{dcolumn}
\usepackage{bm}
\usepackage{bm}
\usepackage{blindtext}
\usepackage{natbib}
\usepackage{booktabs}
\usepackage{mathrsfs}
\usepackage{epstopdf}

\usepackage{textcomp}
\usepackage{times}
\usepackage{float}
\usepackage{latexsym,amsmath,amssymb,bm,euscript}
\usepackage{subfigure}
\usepackage{ulem}

\usepackage{graphicx}
\usepackage{ulem}
\def\be{\begin{equation}}
\def\ee{\end{equation}}
\def\bea{\begin{eqnarray}}
\def\eea{\end{eqnarray}}

\DeclareGraphicsExtensions{.pdf, .png, .jpg, .eps}

\begin{document}

\title{Spinons, solitons and random singlets in the spin-chain compound copper benzoate}

\author{Ying Chen}
\thanks{These authors contributed equally to this study.}
\affiliation{School of Physics and Beijing Key Laboratory of Opto-electronic Functional Materials $\&$ Micro-nano Devices, Renmin University of China, Beijing, 100872, China}

\author{Guijing Duan}
\thanks{These authors contributed equally to this study.}
\affiliation{School of Physics and Beijing Key Laboratory of Opto-electronic Functional Materials $\&$ Micro-nano Devices, Renmin University of China, Beijing, 100872, China}

\author{Yuejiu Zhao}
\thanks{These authors contributed equally to this study.}
\affiliation{Kavli Institute for Theoretical Sciences and CAS Center for Excellence in Topological Quantum Computation,
 University of Chinese Academy of Sciences, Beijing 100190, China}

\author{Ning Xi}
\thanks{These authors contributed equally to this study.}
\affiliation{Institute of Theoretical Physics, Chinese Academy of Sciences, Beijing 100190, China}

\author{Bingying Pan}
\affiliation{College of Physics and Optoelectronic Engineering, Ocean University of China, Qingdao, Shandong 266100, China}

\author{Xiaoyu Xu}
\affiliation{School of Physics and Beijing Key Laboratory of Opto-electronic Functional Materials $\&$ Micro-nano Devices, Renmin University of China, Beijing, 100872, China}

\author{Zhanlong Wu}
\affiliation{School of Physics and Beijing Key Laboratory of Opto-electronic Functional Materials $\&$ Micro-nano Devices, Renmin University of China, Beijing, 100872, China}

\author{Kefan Du}
\affiliation{School of Physics and Beijing Key Laboratory of Opto-electronic Functional Materials $\&$ Micro-nano Devices, Renmin University of China, Beijing, 100872, China}

\author{Shuo Li}
\affiliation{Beijing National Laboratory for Condensed Matter Physics and Institute of Physics, Chinese Academy of Sciences, Beijing, 100190, China}

\author{Ze Hu}
\affiliation{Institute of High Energy Physics, Chinese Academy of Sciences (CAS), Beijing 100049, China}
\affiliation{Spallation Neutron Source Science Center, Dongguan 523803, China}

\author{Rui Bian}
\affiliation{School of Physics and Beijing Key Laboratory of Opto-electronic Functional Materials $\&$ Micro-nano Devices, Renmin University of China, Beijing, 100872, China}

\author{Xiaoqun Wang}
\affiliation{School of Physics, Zhejiang University, Hangzhou, Zhejiang 310027, China}

\author{Wei Li}
\affiliation{Institute of Theoretical Physics, Chinese Academy of Sciences, Beijing 100190, China}

\author{Long Zhang}
\affiliation{Kavli Institute for Theoretical Sciences and CAS Center for Excellence in Topological Quantum Computation,
 University of Chinese Academy of Sciences, Beijing 100190, China}

\author{Yi Cui}
\email{cuiyi@ruc.edu.cn}
\affiliation{School of Physics and Beijing Key Laboratory of Opto-electronic Functional Materials $\&$ Micro-nano Devices, Renmin University of China, Beijing, 100872, China}
\affiliation{Key Laboratory of Quantum State Construction and Manipulation (Ministry of Education), Renmin University of China, Beijing, 100872, China}

\author{Shiyan Li}
\email{shiyan\_li@fudan.edu.cn}
\affiliation{State Key Laboratory of Surface Physics and Department of Physics, Fudan University, Shanghai, 200438, China}

\author{Rong Yu}
\email{rong.yu@ruc.edu.cn}
\affiliation{School of Physics and Beijing Key Laboratory of Opto-electronic Functional Materials $\&$ Micro-nano Devices, Renmin University of China, Beijing, 100872, China}
\affiliation{Key Laboratory of Quantum State Construction and Manipulation (Ministry of Education), Renmin University of China, Beijing, 100872, China}

\author{Weiqiang Yu}
\email{wqyu\_phy@ruc.edu.cn}
\affiliation{School of Physics and Beijing Key Laboratory of Opto-electronic Functional Materials $\&$ Micro-nano Devices, Renmin University of China, Beijing, 100872, China}
\affiliation{Key Laboratory of Quantum State Construction and Manipulation (Ministry of Education), Renmin University of China, Beijing, 100872, China}

\begin{abstract}

The $S=1/2$ antiferromagnetic Heisenberg chain is a paradigmatic quantum system hosting exotic excitations such as spinons and solitons, and forming random singlet state in the presence of quenched disorder. Realizing and distinguishing these excitations in a single material remains a significant challenge. Using nuclear magnetic resonance (NMR) on a high-quality single crystal of copper benzoate, we identify and characterize all three excitation types by tuning the magnetic field at ultra-low temperatures. At a low field of 0.2~T, a temperature-independent spin-lattice relaxation rate ($1/T_1$) over more than a decade confirms the presence of spinons. Below 0.4~K, an additional relaxation channel emerges, characterized by $1/T_1 \propto T$ and a spectral weight growing as $-\ln(T/T_0)$, signaling a random-singlet ground state induced by weak quenched disorder. At fields above 0.5~T, a field-induced spin gap $\Delta \propto H^{2/3}$ observed in both $1/T_1$ and the Knight shift signifies soliton excitations. Our results establish copper benzoate as a unique experimental platform for studying one-dimensional quantum integrability and the interplay of disorder and correlations.
\end{abstract}

\maketitle

\noindent
One-dimensional (1D) quantum spin systems are ideal platforms for exploring emergent phases and novel excitations in condensed matter physics, whose theoretical descriptions are particularly precise. One celebrated example is the $S=1/2$ Heisenberg antiferromagnetic chain (HAFC). Here, the Bethe ansatz solution lays the theoretical foundation for understanding 1D quantum magnets~\cite{1931_EPJA_Bethe}. Subsequently, Faddeev and coworkers later elucidated that the continuous spin excitation spectrum of this system is attributed to unique spin-1/2 fractional excitations, known as spinons~\cite{1981_PLA_Faddeev,1962_PR_Cloizeaux,1981_PRB_muller,1997_PRB_karbach}. More exotic string~\cite{2018_NP_JiandaWu,2019_PRL_JiandaWu,2020_NP_JiandaWu} and E$_8$~\cite{2010_scicence_coldea,2014_PRL_JiandaWu,2021_PRL_cui} excitations have recently been reported in several quasi-1D spin chains with Ising anisotropy, stimulating further research in 1D quantum magnets. In the presence of strong disorder, a random singlet (RS) state is predicted to emerge in 1D systems. It is characterized by the freezing of paired spin singlets over length scales extending to infinity~\cite{1979_PRL_MaDasguptaHu, 1980_PRB_DasguptaMa, 1982_PRL_BhattLee,1992_PRL_BhattFisher,1994_PRB_DSfisher}. The realization and characterization of these novel states and excitations in quasi-1D materials would provide not only confirmation of theoretical predictions in 1D, but also valuable insights for extending novel phases and fractional excitations beyond 1D~\cite{1973_MRB_ANDERSON,2004_PRB_TSenthil}.

Inelastic neutron scattering (INS) studies on the quasi-1D quantum antiferromagnet KCuF$_3$ above the N\'{e}el temperature provided the first confirmation of one-dimensional continuum excitations in the HAFC~\cite{1993_PRL_tennant, 1995_PRB_tennant}.
The quasi-1D material copper benzoate, Cu(C$_6$H$_5$COO)$_2$$\cdot$3H$_2$O, is another ideal system of the $S=1/2$ HAFC~\cite{1963_JPSJ_koizumi}. It consists of Cu$^{2+}$ chains with dominant intrachain exchange coupling~\cite{1970_PTPS_Date,1996_PRB_Dender,2006_PB_wolter}, while the very weak interchain coupling accounts for the N\'{e}el ordering occurring only below 0.8~mK~\cite{2003_PB_Karaki}.
Its specific heat and thermal conductivity above 0.3~K exhibit a linear temperature dependence, consistent with the existence of 1D spinons~\cite{1980_JPSJ_takeda,2022_PRL_LSY}.
However, behaviors inconsistent with a simple spinon description have also been observed: (i) An early NMR study reported a power-law temperature dependence of the spin lattice relaxation rate, $1/T_1 \sim T^{-0.25}$~\cite{1977_JPSJ_ajiro}; (ii) Under an applied magnetic field, an INS study revealed gapped spectral peaks at incommensurate wave vectors~\cite{1981_PRB_muller,1997_PRL_Dender}; and (iii) Thermal conductivity measurements revealed an abrupt downturn below 0.3~K~\cite{2022_PRL_LSY}.

\begin{figure}[t]
\centerline{\includegraphics[width=0.8\columnwidth]{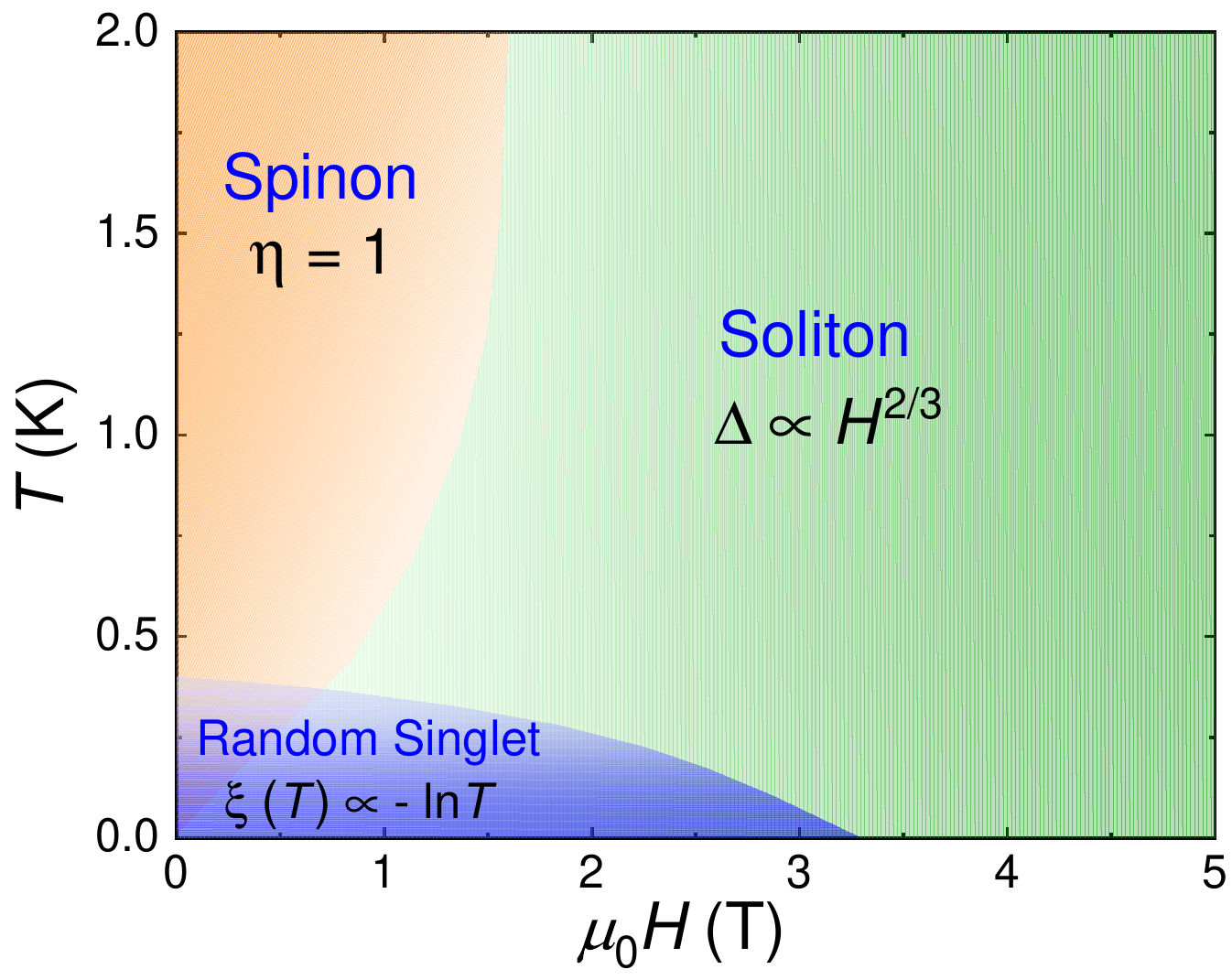}}
\caption{\label{pdsketch} {\bf Schematic $H$-$T$ phase diagram of copper benzoate.} The diagram shows three regimes with different dominant low-energy excitations: spinons, solitons, and RS state. The characteristic scaling behavior in each regime is indicated.
}
\end{figure}

To resolve these issues and reveal the genuine low-energy excitations of the system, we performed an NMR study on a high-quality single crystal of copper benzoate, achieving fields down to 0.2~T and temperatures down to 0.03~K. These extreme conditions proved essential to distinguish different excitation modes, which dominate in three distinct regions of the $H$-$T$ phase diagram as illustrated in Fig.~\ref{pdsketch}.
Under the lowest field of 0.2~T, we observe a nearly temperature-independent $1/T_1$ from 1.4~K down to 0.03~K, which provides strong evidence for spinon excitations.
Applying an inverse Laplace transformation analysis (ILTA) on the nuclear magnetization, we further identified an additional gapless excitation mode, characterized by a spectral weight that increases logarithmically at ultra-low temperatures. This behavior is attributed to an experimental realization of a 1D RS state with correlation length that diverges as $\xi \propto -\ln T$.
At high fields, we observe a large upturn in the nuclear spin-lattice relaxation rate, $K_{\rm n}$, at ultra-low temperatures, which follows a thermal activation behavior with a gap scaling as $\Delta \sim H^{2/3}$. In parallel, a similar gapped behavior is observed in $1/T_1$. These findings are consistent with soliton excitations arising from the field-induced staggered magnetization~\cite{1999_PRB_essler}.
Thus, our work on copper benzoate establishes concrete NMR identification of spinons, solitons, and the long-sought RS state, and demonstrates that their emergence is governed by the interplay of the crystal structure and quenched disorder in the ultra-low-temperature regime.\\

\noindent
{\large \bf Results}

\noindent
{\bf Spectral characterization of the 1D nature}

\begin{figure}[t]
\centerline{\includegraphics[width=\columnwidth]{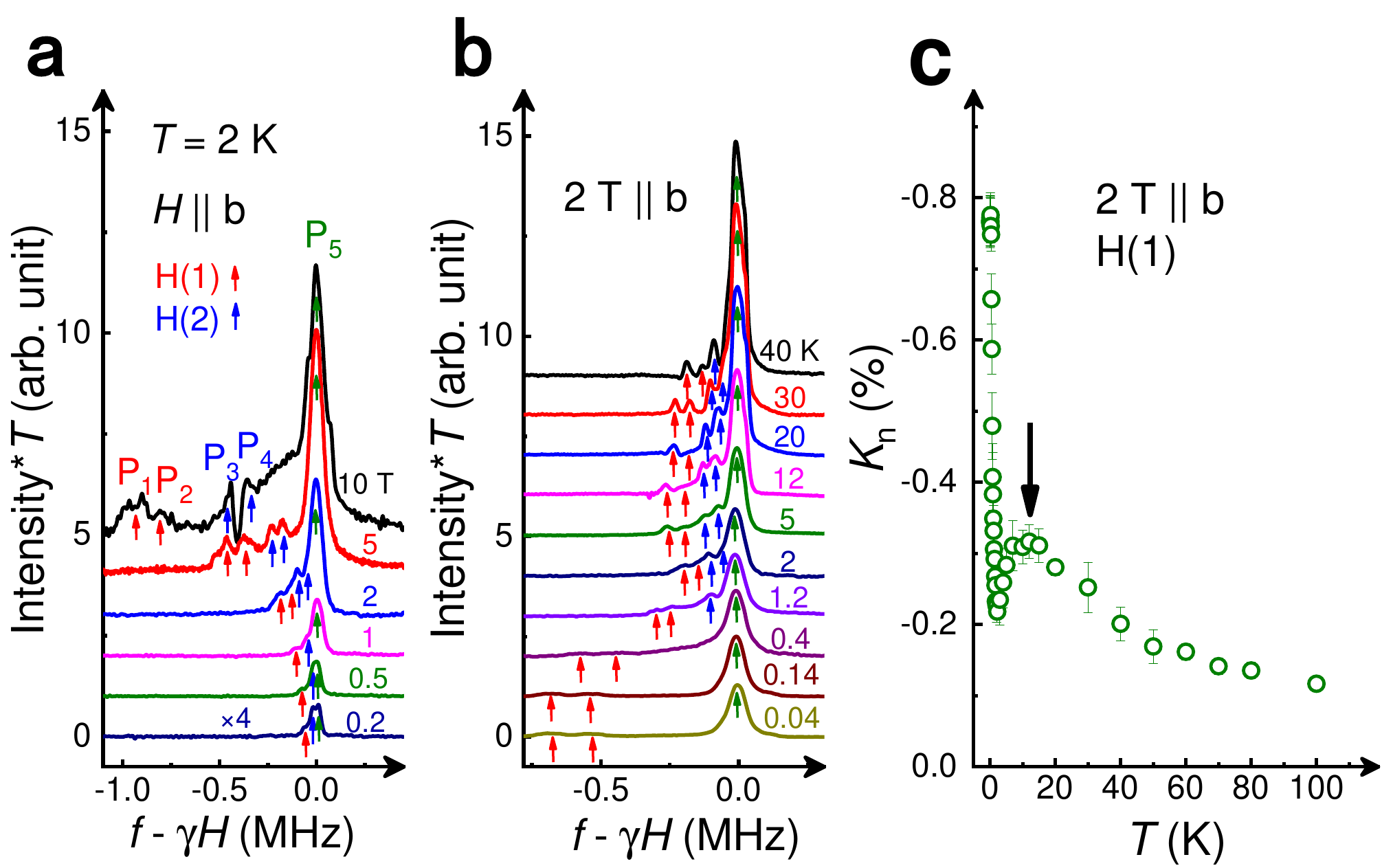}}
\caption{\label{spec}{\bf Field- and temperature-dependent $^1$H NMR response.}
{\bf a} NMR spectra acquired at 2~K with increasing magnetic fields. Peaks are labeled P$_1$–P$_5$. Red and blue arrows indicate peaks assigned to the H(1) and H(2) nuclear sites, respectively; green arrows denote peaks from the H(3) site or benzoate groups.
{\bf b} NMR spectra measured at 2~T with various temperatures down to 0.04~K.
{\bf c} Knight shift $K_{\rm n}$ measured at the H(1) site for magnetic fields of 2~T. The peak position in $K_{\rm n}$ is indicated by the downward arrow.
}
\end{figure}

\noindent
Figure~\ref{spec}a shows the $^1$H NMR spectra obtained at 2~K under different magnetic fields. Based on the crystal structure, which contains six inequivalent protons in water molecules and ten in benzoate groups (Supplementary Fig.~S1), one expects sixteen NMR lines for an arbitrary field orientation, but only four when the field is aligned precisely along the $b$-axis~\cite{1970_PTPS_Date}. Experimentally, we resolve three peaks at 0.2~T and five peaks at 2~T and above, labeled P$_1$–P$_5$ in Fig.~\ref{spec}a. This number of observed peaks indicates that the magnetic field was applied slightly off the crystallographic $b$-axis~\cite{1970_PTPS_Date}. Based on their respective frequency shifts, we assign peaks P$_1$–P$_4$ to the H(1) and H(2) nuclear positions in water molecules, which experience stronger hyperfine couplings. Peak P$_5$ is attributed to either the H(3) site in water molecules or to protons in the benzoate groups, both of which have weaker hyperfine couplings (Supplementary Fig.~S1b).

Figure~\ref{spec}b displays the NMR spectra measured at a fixed field of 2~T as a function of temperature. Down to 0.04~K, no spectral line splitting is observed, indicating the absence of magnetic ordering within the investigated temperature range. The Knight shift $K_{\rm{n}}$ at the H(1) site for field of 2~T was determined from the average frequency of the corresponding paired peaks to improve accuracy, and is plotted as a function of temperature in Fig.~\ref{spec}c.

Upon cooling below 100~K, $K_{\rm{n}}$ increases and develops a peak at approximately 12~K, marked by the downward arrow in Fig.~\ref{spec}c. This peak temperature coincides with that observed in the magnetic susceptibility~\cite{1970_PTPS_Date} and is identified as the Bonner-Fisher peak at $T_{\rm max} = 0.641J$, characteristic of the $S=1/2$ Heisenberg antiferromagnetic chain~\cite{1964_PR_Bonner,1994_PRL_eggert}. Using this relation, the intrachain exchange coupling is estimated as $J \approx 18$~K, consistent with previous reports~\cite{1996_PRB_Dender}. Upon further cooling below 2~K, $K_{\rm{n}}$ exhibits a prominent upturn, which will be discussed below as a signature of field-induced soliton excitations.
\\

\begin{figure*}[t]
\centerline{\includegraphics[width=\textwidth]{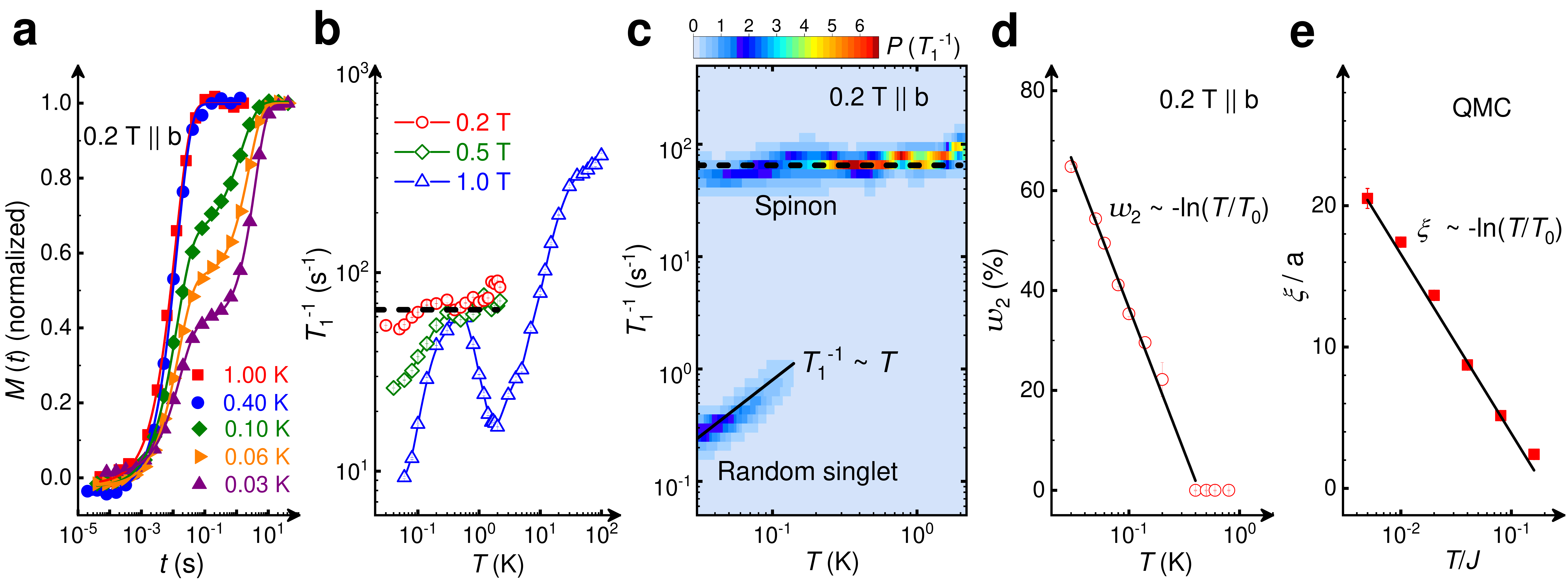}}
\caption{\label{slrr1}{\bf Low-field spin-lattice relaxation rates.}
{\bf a} Nuclear magnetization $M(t)$ as a function of time, measured at the P$_5$ peak under a magnetic field of 0.2~T at various temperatures. Solid lines represent fits to the data using a single-exponential function (at 0.4~K and above) and a double-exponential function (at 0.1~K and below) to extract the spin-lattice relaxation time $T_1$.
{\bf b} The fast component of $1/T_1$ as a function of temperature, measured at different magnetic fields. The dashed line is a guide to the eye, indicating a constant $1/T_1$ at low temperatures.
{\bf c} Spectral distribution $P(1/T_1)$ as a function of temperature obtained from the ILTA of $M(t)$ at 0.2~T. The color scale represents the relative spectral weight. The dotted and solid lines indicate constant $1/T_1$ and $1/T_1 \sim T$ behaviors for the fast and slow components, respectively.
{\bf d} Temperature dependence of the relative weight $w_2$ of the slow $T_1$ component in $M(t)$, derived from $P(1/T_1)$ at 0.2~T. The solid line is a fit to the function $w_2 = -a \ln(T/T_0)$ with $T_0 \approx 0.4$~K.
{\bf e} Correlation length $\xi$ (in units of the lattice constant $a$) as a function of temperature, obtained from quantum Monte Carlo (QMC) simulations for the RS state. The solid line is a fit to $\xi/a \propto -\ln(T/T_0)$.
}
\end{figure*}

\noindent
{\bf Low-field spin-lattice relaxation}

\noindent
To probe spin excitations in the zero-field limit, we carried out $1/T_1$ measurements at a low field of 0.2~T. The P$_5$ site [Fig.~\ref{spec}a] was selected for its weaker hyperfine coupling compared to other nuclear sites, which results in a $T_1$ value that falls within the measurement window.

The nuclear magnetization recovery $M(t)$ at the P$_5$ site is plotted against the time delay $t$ in Fig.~\ref{slrr1}a. Above 0.4~K, the data are well described by a single-exponential function, $M(t) = M(\infty) - a e^{-t/T_1}$, from which $1/T_1$ is extracted. Below 0.1~K, however, the recovery curves clearly resolve both fast and slow relaxation components, requiring a double-exponential fit of the form $M(t) = M(\infty) - a_f e^{-t/T_{1f}} - a_s e^{-t/T_{1s}}$. In the following, we present the fast ($1/T_{1f}$) and slow ($1/T_{1s}$) relaxation components separately.
\\

\noindent
{\bf Fast $T_1$ component and spinon excitations}

\noindent
Figure~\ref{slrr1}b displays the fast component of $1/T_1$, $1/T_{1f}$, plotted as a function of temperature for different magnetic fields. At 0.2~T, $1/T_{1f}$ decreases gradually below 2~K and becomes nearly constant between 1.4~K and 0.03~K. A constant $1/T_1$ over a decade in temperature is a hallmark of spinon excitations in the 1D HAFC~\cite{1994_PRB_sachdev}. This can be understood from the scaling relation $1/T_1 \sim T^{\eta-1}$ for spinons with a linear dispersion $\omega \sim k$, where $\eta$ is the correlation exponent~\cite{1987_PRB_Haldane,1989_PRB_Singh,1990_PRL_Liang}. For the $S=1/2$ HAFC, $\eta = 1$, which directly yields a temperature-independent $1/T_1$~\cite{1987_PRB_Haldane,1989_PRB_Singh,1990_PRL_Liang}. The clear observation of this behavior in our data underscores the high quality of the single crystal and benefits from the local nature of the NMR probe, which, as shown in Fig.~\ref{slrr1}a, segregates the effects of disorder into a separate, slow relaxation channel. We note that $1/T_{1f}$ at 0.2~T shows a weak downturn below 0.3~K. This downturn is enhanced rapidly with increasing magnetic field (evident in the 0.5~T and 1~T data), indicating a field-induced origin, which we attribute to soliton excitations as discussed in the following section.

By contrast, the upturn in $1/T_1$ reported in earlier work~\cite{1977_JPSJ_ajiro} can likely be attributed to the different experimental conditions, such as higher temperatures (above 1.5~K), a higher magnetic field (0.5~T), and the use of a polycrystalline sample.
\\

\begin{figure*}[t]
\centerline{\includegraphics[width=18cm]{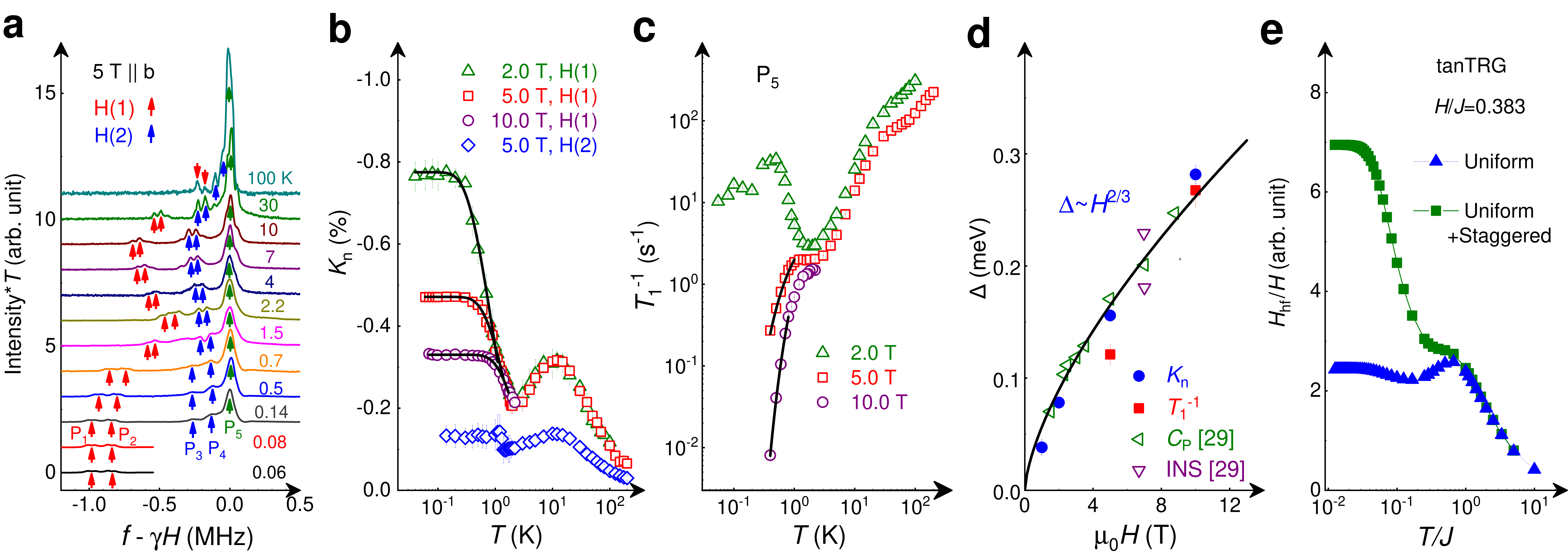}}
\caption{\label{slrr2}{\bf Field-induced gap and soliton excitations.}
{\bf a} NMR spectra measured at 5~T as a function of temperature.
{\bf b} Knight shift $K_{\rm{n}}$ as a function of temperature at selected magnetic fields. Solid lines are fits using a thermal activation form (see text).
{\bf c} Spin-lattice relaxation rate $1/T_1$ at the P$_5$ site as a function of temperature. Solid lines are fits based on a gapped excitation spectrum (see text).
{\bf d} Magnetic field dependence of the excitation gap $\Delta$ extracted from different measurements. Data from specific heat ($C_{\rm P}$) and INS~\cite{1997_PRL_Dender} are included for comparison. The solid line is a fit to $\Delta = 1.85 J (h/J)^{2/3}$.
{\bf e} Dipolar hyperfine field $H_{\rm hf}$ at the H(1) site calculated using the tanTRG method. Solid triangles and squares denote results with and without the staggered magnetization component included in addition to the uniform magnetization.
}
\end{figure*}

\noindent
{\bf Slow $T_1$ component and quenched disorder}

\noindent
To quantitatively resolve the two-component relaxation, we analyzed the recovery curves $M(t)$ using inverse Laplace transformation analysis (ILTA) with Tikhonov regularization~\cite{2002_JMR_song,2020_PRB_singer,2020_NC_Papawassiliou,2021_NP_wangJM} (for details, see Supplementary Sec.~S3). This procedure yields the spectral probability function $P(1/T_1)$, which characterizes the distribution of relaxation rates. In Fig.~\ref{slrr1}c, we present $P(1/T_1)$ at 0.2~T as a colored contour map versus temperature. A corresponding result at 0.5~T is provided in Supplementary Fig.~S3. Thus, $P(1/T_1)$ directly visualizes the distribution of relaxation dynamics in the system.

Above 0.4~K, the spin-lattice relaxation is dominated by a single component with a large, nearly constant $1/T_1$, consistent with the behavior shown in Fig.~\ref{slrr1}b. Below this temperature, a second, slow relaxation channel emerges. The fast component $1/T_{1f}$ remains broadly temperature-independent before showing the field-induced suppression below 0.2~K, as previously discussed. In contrast, the slow component $1/T_{1s}$ is two orders of magnitude smaller and emerges below 0.4~K; its relaxation rate decreases linearly with temperature upon cooling. This clear power-law temperature dependence signals a new type of gapless excitation not captured by the pure HAFC model. As we elaborate below, our analysis identifies this contribution as a direct consequence of quenched disorder.

Taking advantage of the 1D character of the material, we were able to extract the correlation length of this additional gapless phase. The relative spectral weight of the slow $T_1$ component, denoted $w_2$, is calculated from the $P(1/T_1)$ distribution in Fig.~\ref{slrr1}c and plotted as a function of temperature in Fig.~\ref{slrr1}d. Remarkably, over more than a decade in temperature (0.03~K to 0.4~K), $w_2$ is well described by the logarithmic form $w_2 = -a \ln (T/T_0)$, as shown by the solid line fit, yielding the parameters $T_0 = 0.4$~K and $a = 25$. Within the 1D limit, the increase of $w_2$ directly reflects the growth of the thermal correlation length $\xi$. The observed logarithmic dependence, $\xi \propto -\ln(T/T_0)$, is distinct from the conventional quantum critical scaling $\xi \sim T^{-1/z}$. Instead, it signifies that the ground state is a disorder-induced 1D RS state, governed by an infinite-randomness fixed point with an effective dynamical exponent $z \rightarrow \infty$~\cite{1979_PRL_MaDasguptaHu, 1980_PRB_DasguptaMa, 1982_PRL_BhattLee,1992_PRL_BhattFisher,1994_PRB_DSfisher}.

We therefore associate the parameter $T_0$ with the effective energy scale of the disorder and the amplitude $a$ with its effective concentration. From the fitted parameters, we can trace the evolution of the disorder-dominated regions: their effect becomes negligible above ${\sim}0.4$~K, but according to the relation $w_2 = -a \ln(T/T_0)$, the RS state is expected to percolate throughout the entire system ($w_2 \rightarrow 1$) at an extrapolated temperature near 8~mK.

Based on the preceding analysis, we conclude that the high quality of our single crystals allows the separation of two distinct spatial contributions to the relaxation dynamics: one originating from the intrinsic spinon excitations of the pure HAFC, and the other from regions affected by weak quenched disorder. The influence of this disorder manifests below 0.4~K and eventually dominates the low-temperature behavior.

Theoretically, the presence of weak quenched disorder unavoidable in real materials is known to drive the HAFC into a RS state~\cite{1979_PRL_MaDasguptaHu, 1980_PRB_DasguptaMa, 1982_PRL_BhattLee,1992_PRL_BhattFisher,1994_PRB_DSfisher}. This state comprises an ensemble of spin-singlet pairs with a broad distribution of effective exchange couplings in the low-energy limit. This RS scenario predicts quasi-localized spin excitations that yield a thermal conductivity with a quadratic temperature dependence~\cite{2024_PRB_ZhangLong}, thereby offering a consistent explanation for the suppression of low-temperature thermal conductivity observed in copper benzoate~\cite{2022_PRL_LSY}.

The observation of gapless excitations in the present system provides further evidence for the RS state. The formation of long-distance singlets closes the spin gap, resulting in the gapless behavior observed in the slow $1/T_{1s}$ component. Furthermore, the broad distribution of relaxation timescales inherent to the RS state is directly visualized in the spectral distribution of $1/T_1$ around the linear-$T$ trajectory in Fig.~\ref{slrr1}c.

The fact that the disorder effect becomes apparent only below 0.4~K (${\approx}0.02J$) attests to the weakness of the disorder in our compound. This condition is distinct from the high-temperature, strong-disorder limit treated in early theoretical work~\cite{1979_PRL_MaDasguptaHu, 1980_PRB_DasguptaMa, 1982_PRL_BhattLee,1994_PRB_DSfisher,1992_PRL_BhattFisher}. To bridge this gap, we performed quantum Monte Carlo (QMC) simulations of a 1D HAFC with random-bond disorder of strength ${\Delta}J/J \approx 0.2$ (Supplementary Sec.~S4). As shown in Fig.~\ref{slrr1}e, the computed correlation length indeed exhibits a clear $-\ln T$ dependence. This result leads us to a unifying conclusion: despite the difference in disorder strength, the spin dynamics in both the strong- and weak-disorder cases are governed by the same universal logarithmic scaling within their respective temperature regimes.
\\

\noindent
{\bf High-field gapped excitations}

\noindent
Figure~\ref{slrr2}a presents the NMR spectra measured at 5~T across a temperature range from 100~K down to 60~mK. To accurately trace the temperature dependence of the Knight shift $K_{\rm{n}}$, we focused on the H(1) site, leveraging its large hyperfine coupling. The resulting $K_{\rm{n}}$ values measured between 2~T and 10~T are displayed in Fig.~\ref{slrr2}b. A prominent feature observed across these fields is that upon cooling below approximately 2~K, $K_{\rm{n}}$ exhibits a marked upturn and eventually saturates to a constant value at the lowest temperatures, as clearly seen in Fig.~\ref{slrr2}b.

The temperature dependence of $K_{\rm n}$ at each field is well described by a thermal activation function, $K_{\rm n} = a e^{-\Delta / k_{\rm B} T} + b$, where $\Delta$ represents the excitation gap and $b$ the low-temperature saturation value. The corresponding fits are shown as black lines in Fig.~\ref{slrr2}b. The values of $\Delta$ obtained from these fits for fields from 1~T to 10~T are plotted in Fig.~\ref{slrr2}d, revealing a monotonic increase with magnetic field. This analysis confirms that the low-temperature upturn and saturation of $K_{\rm n}$ are characteristic of a thermally activated behavior across the measured field range.

We also measured the spin-lattice relaxation rate $1/T_1$ at the P$_5$ site under high magnetic fields. The fast component of $1/T_1$, measured at different fields, is plotted as a function of temperature in Fig.~\ref{slrr2}c. At fields of 5~T and 10~T, a prominent drop in $1/T_1$ is observed at low temperatures, indicating gap-opening behavior. We then fitted the $1/T_1$ data to a thermal activation function, $1/T_1 = a e^{-\Delta / k_{\rm B} T}$, where $\Delta$ represents the excitation gap.

The values of the gap $\Delta$ extracted from both $K_{\rm{n}}$ and $1/T_1$ are compiled in Fig.~\ref{slrr2}d. For comparison, we also include the gaps reported from specific heat and INS measurements~\cite{1997_PRL_Dender}, which show excellent agreement with our NMR results. Furthermore, the field dependence of the gap is described by the scaling relation $\Delta \propto H^{2/3}$, as evidenced by the solid-line fit in Fig.~\ref{slrr2}d.

The gapped behavior and its field dependence are accounted for by a field-induced effective staggered field~\cite{1997_PRL_oshikawa,1999_PRB_essler,2005_PRL_wolter}. In copper benzoate, the Cu$^{2+}$ ions reside in distorted octahedral coordination environments, resulting in two inequivalent tetragonal axes~\cite{1970_PTPS_Date}. When an external magnetic field is applied along the crystalline $b$-axis, it induces an effective staggered field component along the perpendicular direction~\cite{1970_PTPS_Date}. This staggered field acts as a relevant perturbation to the Luttinger liquid behavior of the HAFC under a uniform field. The corresponding low-energy effective theory is the integrable sine-Gordon model, whose elementary excitations are solitons with a gap scaling as ${\Delta} \propto H^{2/3}$~\cite{1997_PRL_oshikawa,1999_PRB_essler,2005_PRL_wolter}. The precise theoretical form is given by ${\Delta} = 1.85 J (h/J)^{2/3}$~\cite{1997_PRL_Dender}, which is plotted as the solid line in Fig.~\ref{slrr2}d and shows excellent agreement with all experimental data.

To understand the origin of the low-temperature upturn in $K_{\rm n}$, we performed numerical simulations of the relevant Hamiltonian (Supplementary Eq.~S3) using the finite-temperature tangent-space tensor renormalization group (tanTRG) method~\cite{2023_PRL_li,2024_arxive_xi}, with parameters fitted to copper benzoate~\cite{1970_PTPS_Date}. Assuming the hyperfine coupling at the H(1) site is primarily of dipolar form, the hyperfine field $H_{\rm hf}$ includes contributions from both the uniform and the field-induced staggered magnetization of the Cu$^{2+}$ moments. The calculated $H_{\rm hf}$ for a field of $H = 0.383 J$ (corresponding to 5~T) is plotted as a function of temperature in Fig.~\ref{slrr2}e. When only the uniform magnetization is included, a weak low-temperature upturn is present. However, when the staggered magnetization is also considered, a pronounced upturn emerges below ${\sim}0.15 J$ (about 2.7~K), which closely reproduces the experimental data in Fig.~\ref{slrr2}b. This confirms that the low-temperature upturn in $K_{\rm n}$ is dominated by the field-induced staggered magnetization, whose influence diminishes with increasing temperature.
\\

\noindent
{\large \bf Discussion}

\noindent
By performing NMR measurements under very low magnetic fields on a high-quality single crystal, we have unambiguously identified both intrinsic spinon excitations and a disorder-induced RS state at ultra-low temperatures. To our knowledge, our results provide the first direct NMR evidence for the coexistence of these two types of excitations in a 1D quantum magnet, taking advantage of the interplay between crystal structure and quenched disorder at ultra-low temperatures. The spinon excitations are revealed by a nearly constant $1/T_1$ in more than a decade of temperature. 

Experimental evidence for a RS state are directly identified and characterized by a linear temperature dependence of $1/T_1$ and a logarithmic divergence of the correlation length.
The emergence of the RS state readily accounts for the significant suppression of thermal conductivity below 0.3~K reported in this material~\cite{2022_PRL_LSY}: the localized spin excitations within the randomly distributed singlets impede the propagation of itinerant spinons, thereby suppressing spin thermal transport.

We note that the theoretical RS scenario is strictly defined in the zero-field limit, which presents an apparent tension with our experimental identification of the RS state under a finite field of 0.2~T. While an applied field indeed tends to polarize long-distance singlets, it also reduces the triplet excitation gap for spin pairs with an exchange coupling $J$ comparable to the Zeeman energy $g\mu_{\rm B}H$. For those pairs with $\Delta \approx g\mu_{\rm B}H$, the field induces effective gapless excitations. This effect, combined with the suppression of the gap for short-range singlets, ensures a finite low-energy density of states. This explains why the characteristic gapless behavior of the RS state remains consistent with our observations at low magnetic fields.

At high fields, the low-energy gapped solitons are described by an integrable sine-Gordon model derived through
Abelian bosonization~\cite{1999_PRB_essler,1997_PRL_oshikawa}. We note that the coupling constant $\beta$ in this sine-Gordon model varies with the applied field, opening avenues to investigate
other novel excitations predicted by the field theory, such as breathers.
It would also be interesting to connect our findings to the exotic excitations reported in other quasi-1D quantum magnetic systems~\cite{2010_scicence_coldea,2014_PRL_JiandaWu,2021_PRL_cui,2025_PRB_JiandaWu}, which emerge from the cooperative interplay between local crystalline environment and interchain coupling.

This work also demonstrates that NMR serves as an excellent local, low-energy probe for identifying and characterizing novel excitations in one-dimensional systems. The approach established here is directly applicable to the exploration of exotic quantum phases in two-dimensional materials, such as spin liquids~\cite{1973_MRB_ANDERSON, 2021_Kermarrec_PRL} and deconfined quantum critical points~\cite{2004_PRB_TSenthil, 2023_science_cui, 2025_CPL_Cui}.
\\

\noindent
{\bf Materials and methods}

\noindent
Copper benzoate crystallizes in the monoclinic $I2/c$ space group, where the Cu$^{2+}$ ions form 1D chains along the crystalline $c$-axis (see Supplementary Sec.~S1). The single crystal used for NMR measurements was grown by the chemical diffusion method~\cite{2022_PRL_LSY}, with dimensions of 2.5~mm $\times$ 0.4~mm $\times$ 3.0~mm along the $a$-, $b$-, and $c$-axis, respectively. Measurements were performed using a variable-temperature insert (1.7~K and above) and a dilution refrigerator (down to 0.03~K). The magnetic field was applied along the crystalline $b$-axis.

The $^1$H NMR spectra were acquired using the spin-echo method with a typical $\pi/2$-pulse length of approximately $2~\mu$s. At each temperature, the full spectrum was constructed by the frequency sweep. The Knight shift, $K_{\rm n}$, was determined from the relation $K_{\rm n} = (f /\gamma H - 1) \times 100\%$, where $f$ is the resonance frequency of the selected nuclear site, $\gamma = 42.5759$ MHz/T is the proton Zeeman factor, and $H$ is the applied magnetic field. The spin-lattice relaxation time, $T_1$, was measured using the inversion-recovery method. Details on the ILTA of the $T_1$ data and the numerical simulation parameters are provided in Supplementary Secs. S3 and S4.
\\

\noindent
{\bf Data availability}

\noindent
The data that support the findings of this study are available from the corresponding author upon request. 
\\

\noindent
{\bf Competing interests}

\noindent
The authors declare no competing interests.
\\

\noindent
{\bf Acknowledgements}

\noindent
We thank helpful discussions with Profs. Bruce Normand and Qiang Luo.
This work is supported by the National Key Research and Development Program of China (Grant No.~2023YFA1406500),
the National Natural Science Foundation of China (Grants No.~12134020 and No.~12374156),
and Scientific Research Innovation Capability Support Project for Young Faculty (Grant No.~ZYGXQNJSKYCXNLZCXM-M26).
\\

\noindent
{\bf Author contributions.}

\noindent
The project was conceived by W. Y. The samples used in this study were synthesized by Y. Z. and S. L. NMR measurements were carried out by Y. C., X. X., Z. W., K. D., S. L., Z. H., R. B., Y. C., and W. Y. Numerical simulations using the tanTRG method were performed by G.D., N.X., X.W., W.L., and R.Y. QMC calculations were performed by Y. Z. and L. Z. The manuscript was written by Y. C., L. Z., R. Y., and W. Y., with input from all authors. All authors discussed the results and contributed to the final version of the manuscript.

\normalem
\bibliography{cbref}

\end{document}